\documentclass[12pt]{article}

\begin{document}

\title{Horizons and Geodesics of Black Ellipsoids \\
with Anholonomic Conformal Symmetries }
\author{Sergiu I. Vacaru \thanks{%
E-mail address:\ vacaru@fisica.ist.utl.pt, ~~ sergiu$_{-}$vacaru@yahoo.com,\
} \\
{\small \textit{Centro Multidisciplinar de Astrofisica - CENTRA,
Departamento de Fisica,}}\\
{\small \textit{Instituto Superior Tecnico, Av. Rovisco Pais 1, Lisboa,
1049-001, Portugal}}\\
{\small and} \\
Daylan Esmer G\"{o}ksel \thanks{%
E-mail address:\ goksel@istanbul.edu.tr, ~~ gokselde@hotmail.com,\ } \\
{\small \textit{Physics Department, Faculty of Sciences}}\\
{\small \textit{Istanbul University, str. Vezneciler, Instanbul, 34459,
Turkey}}\\
}
\date{May 12, 2003}
\maketitle

\begin{abstract}
The horizon and geodesic structure of static configurations generated by
anisotropic conformal transforms of the Schwarzschild metric is analyzed. We
construct the maximal analytic extension of such off--diagonal vacuum
metrics and conclude that for small deformations there are different classes
of vacuum solutions of the Einstein equations describing "black ellipsoid"
objects \cite{v}. This is possible because, in general, for off--diagonal
metrics with deformed non--spherical symmetries and associated anholonomic
frames the conditions of the uniqueness black hole theorems do not hold. %
\vskip5pt.

Pacs 04.20.Jb,\ 04.70.-s,\ 04.70.Bw

\end{abstract}



\section{Introduction:}

A present research direction is connected to elaboration of new methods of
constructing exact solutions of the Einstein equations parametrized by
off--diagonal metric ansatz and generated by anholonomic transforms and
associated nonlinear connection structures considered on (pseudo) Riemannian
space--times \cite{v,vo}. Such vacuum solutions may be constructed in the
framework of the Einstein gravity theory of arbitrary dimension. They posses
a generic local anisotropy and may describe, for instance, static black hole
and cosmological solutions with ellipsoidal or toroidal symmetry, various
soliton--dilaton configurations and wormholes and flux tubes with
anisotropic polarizations and/or running constants with different extensions
to backgrounds of rotation ellipsoids, elliptic cylinders, bipolar and
toroidal symmetry and anisotropy. The investigation of physical properties
of new classes of solutions presents a substantial interest.

The aim of this paper is to prove that in the framework of general
relativity theory there are static solutions with deformed spherical
symmetries (e. g. to the symmetry of resolution ellipsoid) which may
describe different types of black hole like objects with non--spherical
horizons. In Ref. \cite{vo} we analyzed the horizon and geodesic structure
of a class of static black ellipsoid solutions and concluded that for some
parametrizations and small eccentricity they describe classes of
Schwarschild like vacuum metrics with ''slightly'' deformed horizons. Far a
way from the horizons, such generic off--diagonal metrics with ellipsoid
symmetry may transform into some diagonal ones with spherical symmetry which
results in a compatibility with the asymptotic Minkowski space-time.

For the vacuum exact solutions with spherical symmetry, there were proved
the uniqueness black hole theorems (UBHT) \cite{ut} which define the basic
properties of static and/or stationary gravitational configurations. Such
objects are described by metrics which are diagonal or may be diagonalized
with respect to a local coordinate coordinate frame, $e_{\alpha }=\partial
_{\alpha }=\partial /\partial u^{\alpha },$ satisfying the holonomy relations%
\[
\partial _{\alpha }\partial _{\beta }-\partial _{\beta }\partial _{\alpha
}=0,
\]%
where $u^{\alpha }=\left( x^{i},y^{a}\right) $ are local coordinates on a
four dimensional pseudo--Riemannian space--time $V^{3+1}.$ The new classes
of static and stationary solutions of the vacuum Einstein equations with
ellipsoid and/or toroidal symmetry \cite{v,vth} are parametrized by
off--diagonal metric ansatz which can be diagonalized only with respect to
anholonomic frames
\begin{equation}
e_{\alpha }=A_{\alpha }^{\beta }\left( u^{\gamma }\right) \partial _{\beta },
\label{transftet}
\end{equation}%
subjected to anholonomy \ relations
\begin{equation}
e_{\alpha }e_{\beta }-e_{\beta }e_{\alpha }=W_{\alpha \beta }^{\gamma
}\left( u^{\varepsilon }\right) e_{\gamma },  \label{anhol}
\end{equation}%
where $W_{\alpha \beta }^{\gamma }\left( u^{\varepsilon }\right) $ are
called the coefficients of anholonomy. We illustrated that a very large
class of off--diagonal metrics with coefficients depending on two and three
variables may be diagonalized by corresponding anholonomic transforms (\ref%
{transftet}) and proved that the vacuum Einstein equations (as well with
matter fields for a very general type of energy--momentum tensors) with
anholonomic variables admit exact solutions. Such metrics posses a generic
anisotropy being constructed in the framework of general gravity or
extra/lower dimension Einstein gravity theories. The new type of exact
solutions depends anisotropically on a coordinate $y^{3}=v$ (for static or
stationary configurations chosen as an angular coordinate) and can be
generated by some inter--related anholonomic and conforma transforms of
usual diagonal locally isotropic solutions (for instance, by some
deformations of the Scwarzschild or Reissner--Nordstrom solutions).

We emphasize that for generic off--diagonal metrics and anholonomic
gravitational systems the conditions and proofs of UBHT (if such ones can be
formulated for some particular anisotropic configurations) have to be
revised. We base our arguments on consequences of the Kaluza--Klein gravity:
It is known that for some particular off--diagonal metric ansatz and
corresponding re--definitions of Lagrangians we can model a class of
effective gravitational (with diagonal metrics) and matter fields
(electromagnetic ones or, for higher than five dimensions, \ non--Abelian
gauge fields) interactions. Similar conclusions hold true if there are
considered three dimensional gravitational and matter field interactions
modeled effectively in the framework of four dimensional gravity. It is
obvious that in this case we are not restricted by the conditions of UBHT
because we have additional matter field interactions induced from extra
dimensions or by off--diagonal terms of the metric in the same dimension. Of
course, in general, the vacuum gravitational dynamics of off--diagonal
metric coefficients can not be associated to any matter field contributions;
the may be of geometric, vacuum gravitational nature, when some degrees of
gravitational dynamics are anholonomic (constrained). But, even for such
configurations, we have to revise the conditions of application of UBHT if
the off--diagonal metrics, anholonomic frames and non--spherical symmetries
are considered. The extended non--diagonal dynamics of vacuum gravitational
fields subjected to some anholonomy conditions is very different from that
analyzed in the cases of the well known Scwarzschild, Kerr--Newmann and
Reissner--Nordstrom black hole solutions.

The geometry of pseudo-Riemannian space--times with off--diagonal metrics
may be equivalently modeled by some diagonalized metrics with respect to
anholnomic frames with associated nonlinear connection structure (see
details in Refs. \cite{v,vth}; for nonlinear connections in vector and
spinor bundles and in superbundles the formalism was developed in Ref. \cite%
{ma}). In this case we obtain an effective torsion (a ''pure'' frame effect
vanishing with respect to holonomic frames) which together with the
anholonomy coefficients modify the formulas for the Levi Civita connection
and the corresponding Riemann, Ricci and Einstein tensors by elongating them
with additional terms. As a result, for such configurations we can not prove
the UBHT. That why it was possible to construct in Refs. \cite{v,vth}
various classes of off--diagonal exact vacuum solutions with the same
symmetry of the metric coefficients (for instance, ellipsoid or toroidal
configurations) and to suggest that some of them can describe certain static
black hole objects with non--spherical symmetries which was confirmed by an
analysis of geodesic congruences and of horizons in Ref. \cite{vo}.

In this work, we show that there are alternative possibilities to construct
''deformed'' black holes by considering conformal transforms adapted to the
nonlinear connection structure. We analyze the maximal analytic extension
and the Penrose diagrams of such metrics and state the basic properties by
comparing them with the Reissner--Nordstrom solution.

The paper has the following structure: in section 2 we present the necessary
formulas on off--diagonal metrics with conformal factors which can be
diagonalized with respect to anholonomic frames with associated nonlinear
connection structure and write down the vacuum Einstein equations. In
section 3 we define a class of static anholonomic conformal transforms of
the Scwarzschild metric to some off--diagonal metrics. In section 4 we
construct the maximal analytic extension of such metrics, for static small
ellipsoid deformations, analyze their horizon structure. Section 5 contains
a study of the conditions when the geodesic behaviour of ellipsoidal metrics
can be congruent to the conformal transform of the Schwarzschild one. A
discussion and conclusions are considered in section 6.

\section{Off--diagonal Metrics and Anholonomic Conformal Transforms}

\bigskip\ Let us consider the quadratic line element
\begin{equation}
ds^{2}=\Omega \left( x^{i},v\right) {\hat{g}}_{\alpha \beta }\left(
x^{i},v\right) du^{\alpha }du^{\beta },  \label{metricaux1}
\end{equation}%
with ${\hat{g}}_{\alpha \beta }$ parametrized by the ansatz%
\begin{equation}
\left[
\begin{array}{cccc}
g_{1}+w_{1}^{\ 2}h_{3}+n_{1}^{\ 2}h_{4} & w_{1}w_{2}h_{3}+n_{1}n_{2}h_{4} &
w_{1}h_{3} & n_{1}h_{4} \\
w_{1}w_{2}h_{3}+n_{1}n_{2}h_{4} & g_{2}+w_{2}^{\ 2}h_{3}+n_{2}^{\ 2}h_{4} &
w_{2}h_{3} & n_{2}h_{4} \\
w_{1}h_{3} & w_{2}h_{3} & h_{3} & 0 \\
n_{1}h_{4} & n_{2}h_{4} & 0 & h_{4}%
\end{array}%
\right] ,  \label{ansatz2}
\end{equation}%
for a 4D pseudo--Riemannian space--time $V^{3+1}$ enabled with local
coordinates $u^{\alpha }=\left( x^{i},y^{a}\right) $ where the indices of
type $i,j,k,...$ run values $1$ and $2$ and the indices $a,b,c,...$ take
values $3$ and $4;$ $\ y^{3}=v=\varphi $ and $y^{4}=t$ are considered
respectively as the ''anisotropic'' and time like coordinates. The
components $g_{i}=g_{i}\left( x^{i}\right) ,h_{a}=h_{ai}\left(
x^{k},v\right) $ and $n_{i}=n_{i}\left( x^{k},v\right) $ \ are some
functions of necessary smoothly class or even singular in some points and
finite regions. So, the $g_{i}$--components of our ansatz depend only on
''holonomic'' variables $x^{i}$ and the rest of coefficients may also depend
on one ''anisotropic'' (anholonomic) variable $y^{3}=v;$ the ansatz does not
depend on the time variable $y^{4}=t:$ we shall search for static solutions.

\ The element (\ref{metricaux1}) can be diagonalized,
\begin{equation}
ds^{2}=\Omega \left( x^{i},v\right) \left[ {g}_{1}\left( dx^{1}\right) ^{2}+{%
g}_{2}\left( dx^{2}\right) ^{2}+h_{3}\left( \delta v\right) ^{2}+h_{4}\left(
\delta y^{4}\right) ^{2}\right] ,  \label{dmetr2}
\end{equation}%
with respect to the anholomic co--frame
\begin{equation}
\delta ^{\alpha }=(d^{i}=dx^{i},\delta ^{a}=dy^{a}+N_{i}^{a}dx^{i})=\left(
d^{i},\delta v=dv+w_{i}dx^{i},\delta y^{4}=dy^{4}+n_{i}dx^{i}\right)
\label{ddif4}
\end{equation}%
which is dual to the frame
\begin{equation}
\delta _{\alpha }=(\delta _{i}=\partial _{i}-N_{i}^{a}\partial _{a},\partial
_{b})=\left( \delta _{i}=\partial _{i}-w_{i}\partial _{3}-n_{i}\partial
_{4},\partial _{3},\partial _{4}\right) ,  \label{dder4}
\end{equation}%
where $\partial _{i}=\partial /\partial x^{i}$ and $\partial _{b}=\partial
/\partial y^{b}$ are usual partial derivatives. The tetrads $\delta _{\alpha
}$ and $\delta ^{\alpha }$ are anholonomic because, in general, they satisfy
some anholonomy relations (\ref{anhol}) with non--trivial coefficients%
\[
W_{ij}^{a}=\delta _{i}N_{j}^{a}-\delta
_{j}N_{i}^{a},~W_{ia}^{b}=-~W_{ai}^{b}=\partial _{a}N_{i}^{b}.
\]%
The anholonomy is induced by the coefficients $N_{i}^{3}=w_{i}$ and $%
N_{i}^{4}=n_{i}$ which ''elongate'' partial derivatives and differentials if
we are working with respect to anholonomic frames. \ They define a nonlinear
connection (in brief, N--connection) associtated to an anholonomic frame
structure with mixed holonomic $\left( x^{i}\right) $ and anholonomic $%
\left( y^{a}\right) $ coordinates, see details in Refs. \cite{v,vth,ma}.
Here we emphasize that on (pseudo) Riemannian spaces the N--connection is an
object completely defined by anholonomic frames, when the coefficients of
tetradic transform (\ref{transftet}), $A_{\alpha }^{\beta }\left( u^{\gamma
}\right) ,$ are parametrized explicitly via values $\left( N_{i}^{a},\delta
_{i}^{j},\delta _{b}^{a}\right) ,$ where $\delta _{i}^{j}$ $\ $and $\delta
_{b}^{a}$ are the Kronecker symbols.

By straightforward calculation with respect to the frames (\ref{ddif4})--(%
\ref{dder4}) \cite{vth,v,vo} \ we find the non--trivial components of the
Ricci tensor and of the vacuum Einstein equations,
\begin{eqnarray}
R_{1}^{1}=R_{2}^{2}=-\frac{1}{2g_{1}g_{2}}[g_{2}^{\bullet \bullet }-\frac{%
g_{1}^{\bullet }g_{2}^{\bullet }}{2g_{1}}-\frac{(g_{2}^{\bullet })^{2}}{%
2g_{2}}+g_{1}^{^{\prime \prime }}-\frac{g_{1}^{^{\prime }}g_{2}^{^{\prime }}%
}{2g_{2}}-\frac{(g_{1}^{^{\prime }})^{2}}{2g_{1}}] &=&0,  \label{ricci1a} \\
R_{3}^{3}=R_{4}^{4}=-\frac{\beta }{2h_{3}h_{4}}=-\frac{1}{2h_{3}h_{4}}\left[
h_{4}^{\ast \ast }-h_{4}^{\ast }\left( \ln \sqrt{|h_{3}h_{4}|}\right) ^{\ast
}\right] &=&0,  \label{ricci2a} \\
R_{3i}=-w_{i}\frac{\beta }{2h_{4}}-\frac{\alpha _{i}}{2h_{4}} &=&0,
\label{ricci3a} \\
R_{4i}=-\frac{h_{4}}{2h_{3}}\left[ n_{i}^{\ast \ast }+\gamma n_{i}^{\ast }%
\right] &=&0,  \label{ricci4a}
\end{eqnarray}%
for
\begin{equation}
\delta _{i}h_{3}=0  \label{d1}
\end{equation}%
and
\begin{equation}
\delta _{i}\Omega =0  \label{d2}
\end{equation}%
where
\begin{equation}
\alpha _{i}=\partial _{i}h_{4}^{\ast }-h_{4}^{\ast }\partial _{i}\ln \sqrt{%
|h_{3}h_{4}|},~\gamma =3h_{4}^{\ast }/2h_{4}-h_{3}^{\ast }/h_{3},
\label{abc}
\end{equation}%
and the partial derivatives are denoted $g_{1}^{\bullet }=\partial
g_{1}/\partial x^{1},g_{1}^{^{\prime }}=\partial g_{1}/\partial x^{2}$ and $%
h_{3}^{\ast }=\partial h_{3}/\partial v.$ We imposed additionally the
condition $\delta _{i}N_{j}^{a}=\delta _{j}N_{i}^{a}$ in order to work with
the Levi Civita connection (see a discussion in \cite{vo}) which may be
satisfied, for instance, if
\begin{equation}
w_{1}=w_{1}\left( x^{1},v\right) ,n_{1}=n_{1}\left( x^{1},v\right)
,w_{2}=n_{2}=0,  \label{cond1}
\end{equation}%
or, inversely, if
\[
w_{1}=n_{1}=0,w_{2}=w_{2}\left( x^{2},v\right) ,n_{2}=n_{2}\left(
x^{2},v\right) .
\]%
We note that the condition (\ref{d2}) relates the conformal factor $\Omega $
with the coefficients of the N--connection and with another off--diagonal
metric components, which are also subjected to the conditions (\ref{d1}).
This induces a constrained type, aholonomic (anisotropic), of conformal
symmetry.

The system of equations (\ref{ricci1a})--(\ref{ricci4a}) and the conditions (%
\ref{d1})--(\ref{d2}) can be integrated in general form \cite{vth}. Physical
solutions are defined following some additional boundary conditions, imposed
types of symmetries, nonlinearities and singular behaviour and compatibility
in the locally anisotropic limits with some well known exact solutions.

\section{Anholonomic Conformal Deformations of \newline
the Schwarzschild Metric}

\bigskip We consider a particular parametrization of metrics with
anisotropic conformal factors and non--spherical (in particular,
ellipsoidal) symmetry defined in Refs \cite{vth,v},
\begin{eqnarray}
\delta s^{2} &=&\Omega \left( r,\varphi \right) [-\left( 1-\frac{2m}{r}+%
\frac{\varepsilon }{r^{2}}\right) ^{-1}dr^{2}-r^{2}q(r)d\theta ^{2}
\label{sch} \\
&&-\eta _{3}\left( r,\theta ,\varphi \right) r^{2}\sin ^{2}\theta \delta
\varphi ^{2}+\left( 1-\frac{2m}{r}+\frac{\varepsilon }{r^{2}}\right) \delta
t^{2}]  \nonumber
\end{eqnarray}%
where
\begin{eqnarray*}
\Omega  &=&1+\varepsilon \lambda \left( r,\varphi \right) +\varepsilon
^{2}\gamma \left( r,\varphi \right) , \\
\delta \varphi  &=&d\varphi +w_{1}\left( r,\varphi \right) dr\mbox{ and
}\delta t=dt+n_{1}\left( r,\varphi \right) dr,
\end{eqnarray*}%
for $w_{1}$ and $n_{1}\sim \varepsilon .$ In this paper we shall
not work with exact solutions \ but, for simplicity, we shall
consider some conformal and anholonomic deformations of the
Schwarzschild metric up to the second order on $\varepsilon ,$
considered as a small parameter $\left( 0\leq \varepsilon \ll
1\right) $, which are contained in those exact solutions. The
functions $q(r),\lambda \left( r,\varphi \right) ,$ $\gamma \left(
r,\varphi \right) ,$ $\eta _{3}\left( r,\varphi \right) ,$
$w_{1}\left( r,\varphi \right) $ and $n_{1}\left( r,\varphi
\right) $ will be found as the metric (\ref{sch}) would define a
solution of the vacuum Einstein equations (in the limit
$\varepsilon \rightarrow 0$ and $\eta _{3}\rightarrow 1$ we have
just the Schwarzschild solution for a point particle of mass $m).$

We note that the condition of vanishing of the metric coefficient before $%
\delta t^{2},$
\begin{equation}
\left( 1+\varepsilon \lambda +\varepsilon ^{2}\gamma \right) \left( 1-\frac{%
2m}{r}+\frac{\varepsilon }{r^{2}}\right) =1-\frac{2m}{r}+\varepsilon \frac{%
\Phi }{r^{2}}+\varepsilon ^{2}\Theta =0,  \label{horc}
\end{equation}%
where
\begin{eqnarray*}
\Phi &=&\lambda \left( r^{2}-2mr\right) +1, \\
\Theta &=&\gamma \left( 1-\frac{2m}{r}\right) +\lambda
\end{eqnarray*}%
defines, for instance, a rotation ellipsoid configuration if $\lambda $ and $%
\gamma $ are chosen
\begin{eqnarray}
\lambda &=&\left( 1-\frac{2m}{r}\right) ^{-1}\left( \cos \varphi -\frac{1}{r}%
\right) ,  \nonumber \\
\gamma &=&-\lambda \left( 1-\frac{2m}{r}\right) ^{-1}.  \label{valc}
\end{eqnarray}%
We note that all computations should be performed up to the second order on $%
\varepsilon ,$ but because we may fix some conditions on $\lambda $ and $%
\gamma $ as to make zero the coefficient defining deformations of horizons
which are proportional to $\varepsilon ^{2},$ we may restrict our further
analysis only to linear on $\varepsilon $ deformations.

A zero value of the coefficient (\ref{horc}), with ellipsoid like symmetry
stated by the values (\ref{valc}), in the first order on $\varepsilon ,$
follows \ if
\begin{equation}
r_{+}=\frac{2m}{1+\varepsilon \cos \varphi }=2m[1-\varepsilon \cos \varphi ],
\label{ebh}
\end{equation}%
which is the equation for a 3D ellipsoid like hypersurface with a small
eccentricity $\varepsilon .$ In general, we can consider any arbitrary
functions $\lambda (r,\theta ,\varphi )$ and $\eta _{3}(r,\theta ,\varphi ).$

Having introduced a new radial coordinate
\begin{equation}
\xi =\int dr\sqrt{|1-\frac{2m}{r}+\frac{\varepsilon }{r^{2}}|}  \label{int2}
\end{equation}%
and defined
\begin{eqnarray}
h_{3} &=&-\eta _{3}(\xi ,\theta ,\varphi )r^{2}(\xi )\sin ^{2}\theta ,
\nonumber \\
h_{4} &=&1-\frac{2m}{r}+\frac{\varepsilon }{r^{2}},  \label{sch1q}
\end{eqnarray}%
for $r=r\left( \xi \right) $ found as the inverse function after integration
in (\ref{int2}), we write the metric (\ref{sch}) in coordinates $\left( \xi
,\theta ,\varphi ,t\right) ,$
\begin{eqnarray}
ds^{2} &=&\Omega \left( \xi ,\theta ,\varphi \right) \left[ -d\xi
^{2}-r^{2}\left( \xi \right) q(\xi )d\theta ^{2}+h_{3}\left( \xi ,\theta
,\varphi \right) \delta \varphi ^{2}+h_{4}\left( \xi \right) \delta t^{2}%
\right] ,  \label{sch1} \\
\delta \varphi  &=&d\varphi +w_{1}\left( \xi ,\varphi \right) d\xi
\mbox{
and }\delta t=dt+n_{1}\left( \xi ,\varphi \right) d\xi ,\quad r^{2}\left(
\xi \right) q(\xi )=\xi ^{2},  \nonumber
\end{eqnarray}%
where the coefficient $n_{1}$ is taken to satisfy (\ref{cond1}). $\ $

Let us find the conditions when the coefficients of the metric (\ref{sch})
define solutions of the vacuum Einstein equations. For $%
g_{1}=-1,g_{2}=-r^{2}\left( \xi \right) q(\xi )$ and arbitrary $h_{3}(\xi
,\theta ,\varphi )$ and $h_{4}\left( \xi \right) $ solve the equations (\ref%
{ricci1a})--(\ref{ricci3a}). The coefficients $n_{i}$ are solutions of the
equations (\ref{ricci4a}) with $h_{4}^{\ast }=0,$
\begin{equation}
n_{k}=n_{k[1]}\left( x^{i}\right) +n_{k[2]}\left( x^{i}\right) \int \eta
_{3}(\xi ,\theta ,\varphi )d\varphi ,~h_{3}^{\ast }\neq 0;  \nonumber
\end{equation}%
the functions $n_{k[1,2]}\left( x^{i}\right) $ are to be stated by boundary
conditions. In order to consider linear infinitesimal extensions on $%
\varepsilon $ of the Schwarzschild metric we write
\[
n_{1}=\varepsilon \widehat{n}_{1}\left( \xi ,\varphi \right)
\]%
where
\begin{equation}
\widehat{n}_{1}\left( \xi ,\varphi \right) =n_{1[1]}\left( \xi \right)
+n_{1[2]}\left( \xi \right) \int \eta _{3}\left( \xi ,\varphi \right)
d\varphi .  \label{auxf4}
\end{equation}%
The last step is to find such $\Omega \left( \xi ,\varphi \right) ,$ $w_{i}$
and $\eta _{3}\left( \xi ,\theta ,\varphi \right) $ as to satisfy (\ref{d1}%
)--(\ref{d2}). In the first order on $\varepsilon $ those equations may be
solved by
\[
\Omega =1+\varepsilon h_{3}
\]%
which imposes the condition%
\[
-\eta _{3}\left( \xi ,\theta ,\varphi \right) r^{2}\sin ^{2}\theta =\lambda
\left( \xi ,\varphi \right) +\varepsilon \gamma \left( \xi ,\varphi \right) ,
\]%
and by
\begin{eqnarray*}
w_{i} &=&\partial _{i}\Omega /\Omega ^{\ast },\eta _{3}^{\ast }\neq 0, \\
&=&0,\eta _{3}^{\ast }=0.
\end{eqnarray*}

The data
\begin{eqnarray}
g_{1} &=&-1,g_{2}=-r^{2}(\xi ),\Omega =1+\varepsilon \lambda \left(
r,\varphi \right) +\varepsilon ^{2}\gamma \left( r,\varphi \right) ,
\label{data} \\
h_{3} &=&-\eta _{3}(\xi ,\theta ,\varphi )r^{2}(\xi )\sin ^{2}\theta
,h_{4}=1-\frac{2m}{r(\xi )}+\frac{\varepsilon }{r^{2}(\xi )},  \nonumber \\
w_{1} &=&\partial _{1}\Omega /\Omega ^{\ast },n_{1}=\varepsilon \lbrack
n_{1[1]}\left( \xi \right) +n_{1[2]}\left( \xi \right) \int \eta _{3}\left(
\xi ,\theta ,\varphi \right) d\varphi ],  \nonumber \\
w_{2} &=&0,n_{2}=0,  \nonumber
\end{eqnarray}%
for the metric (\ref{sch}) define a class of solutions of the vacuum
Einstein equations depending on arbitrary functions $\eta _{3}(\xi ,\varphi
),n_{1[1]}\left( \xi \right) $ and $n_{1[2]}\left( \xi \right) $ which
should be defined by some boundary conditions. Such solutions are generated
by small deformations (in particular cases of rotation ellipsoid symmetry)
of the Schwarzschild metric.

In order to connect our solutions with some small deformations of the
Schwar\-zschild metric, as well to satisfy the asymptotically flat
condition, we must chose such functions $n_{k[1,2]}\left( x^{i}\right) $ as $%
n_{k}\rightarrow 0$ for $\varepsilon \rightarrow 0$ and $\eta
_{3}\rightarrow 1.$ These functions have to be selected as to vanish far
away from the horizon, for instance, like $\sim 1/r^{1+\tau },\tau >0,$ for
long distances $r\rightarrow \infty .$

\section{ Analytic Extensions of Conformal Ellipsoid Metrics}

We emphasize that the metric (\ref{sch}) (equivalently (\ref{sch1}))
considered with respect to the \ anholonomic basis has a number of
similarities with the Schwrzschild and Reissner--Nordstrem solutions. We can
identify $\varepsilon $ with $e^{2}$ and treat it as a stationary metric
with effective ''electric'' charge induced by small off--diagonal metric
extensions. We analyzed such metrics in Ref. \cite{vo} where the conformal
factor was considered for a trivial value $\Omega =1$ (the off--diagonal
extension being driven by a different equation). The coefficients of such
metrics are similar to those from the Reissner--Nordstrem solution but given
with respect to a corresponding anholonomic frame and with additional
dependencies on ''polarization functions'' $\eta _{3},n_{1[1,2]}\left( \xi
\right) ,\lambda $ and $\gamma .$ Nevertheless, there is a substantial
difference: the static deformed metric generated by anholnomic transforms
satisfies a solution of the vacuum Einstein equations which differs
substantially from the Reissner--Nordstrem metric being an exact static
solution of the Einstein--Maxwell equations.

For $\varepsilon \rightarrow 0$ and $\eta _{3}\rightarrow 1$ the metric (\ref%
{sch}), as well those considered in Ref. \cite{vo}, transforms into the
usual Schwarzschild metric. We can consider a special type of deformations
which induces ellipsoid symmetries. Such symmetries are determined by
corresponding conditions of vanishing of the coefficient before $\delta t$ $%
\ $which describe ellipsoidal hypersurfaces like for the rotating Kerr
metric, but in our case for a static off--diagonal metric.

In the first order on $\varepsilon $ the metric (\ref{sch}) has
singularities at
\[
r_{\pm }=m\left[ 1\pm \left( 1-\varepsilon \frac{\Phi \left( \varphi \right)
}{2m^{2}}\right) \right] ,
\]%
for $\Phi \left( \varphi \right) $ stated for $r=2m.$ For simplicity, we may
take explicit equations for the hypersurfaces $r_{+}=r_{+}\left( \varphi
\right) $ and analyze the horizon properties for some fixed ''directions''
given in a smooth vecinity of any value $\varphi =\varphi _{0}$ and $%
r_{+}^{0}=r_{+}\left( \varphi _{0}\right) .$ The metrics (\ref{sch}) and (%
\ref{sch1}) are regular in the regions I ($\infty >r>r_{+}^{0}),$ II ($%
r_{+}^{0}>r>r_{-}^{0})$ and III$\;(r_{-}^{0}>r>0).$ As in the Schwarzschild,
Reissner--Nordstrem and Kerr cases these singularities can be removed by
introducing suitable coordinates and extending the manifold to obtain a
maximal analytic extension \cite{gb,carter}. We have similar regions as in
the Reissner--Nordstrem space--time, but with just only one possibility $%
\varepsilon <1$ instead of three relations for static electro--vacuum cases (%
$e^{2}<m^{2},e^{2}=m^{2},e^{2}>m^{2};$ where $e$ and $m$ are correspondingly
the electric charge and mass of the point particle in the
Reissner--Nordstrem metric). Saw we may consider the usual Penrose's
diagrams as for a particular case of the Reissner--Nordstrem space--time but
keeping in mind that such diagrams and horizons may be drawn for a fixed
value of the anisotropic angular coordinate.

We may construct the maximally extended manifold in a similar manner as for
the anisotropic metrics with trivial conformal factors \cite{vo} and to
proceed in steps analogous to those in the Schwarzschild case (see details,
for instance, in Ref. \cite{haw})). At the first step we introduce a new
coordinate
\[
r^{\Vert }=\int dr\left( 1-\frac{2m}{r}+\frac{\varepsilon }{r^{2}}\right)
^{-1}
\]%
for $r>r_{+}^{1}$ and find explicitly the coordinate
\begin{equation}
r^{\Vert }=r+\frac{(r_{+}^{1})^{2}}{r_{+}^{1}-r_{-}^{1}}\ln (r-r_{+}^{1})-%
\frac{(r_{-}^{1})^{2}}{r_{+}^{1}-r_{-}^{1}}\ln (r-r_{-}^{1}),  \label{r1}
\end{equation}%
where $r_{\pm }^{1}=r_{\pm }^{0}$ with $\Phi =1.$ If the coordinate $r$ is
expressed as a function on variable $\xi ,$ than the coordinate $r^{\Vert }$
can be also expressed as a function on $\xi $ depending additionally on some
parameters.

At the second step one defines the advanced and retarded coordinates, $%
v=t+r^{\Vert }$ and $w=t-r^{\Vert },$ with corresponding elongated
differentials
\[
\delta v=\delta t+dr^{\Vert }\mbox{ and }\delta w=\delta t-dr^{\Vert }
\]%
which transform the metric (\ref{sch1}) as%
\[
\delta s^{2}=\Omega \left( \xi ,\theta ,\varphi \right) [-r^{2}(\xi )q(\xi
)d\theta ^{2}-\eta _{3}(\xi ,\theta ,\varphi )r^{2}(\xi )\sin ^{2}\theta
\delta \varphi ^{2}+(1-\frac{2m}{r(\xi )}+\frac{\varepsilon }{r^{2}(\xi )}%
)\delta v\delta w],
\]%
where (in general, in non--explicit form) $r(\xi )$ is a function of type $%
r(\xi )=r\left( r^{\Vert }\right) =$ $r\left( v,w\right) .$ The final steps
consist in introducing some new coordinates $(v^{\prime \prime },w^{\prime
\prime })$ by%
\[
v^{\prime \prime }=\arctan \left[ \exp \left( \frac{r_{+}^{1}-r_{-}^{1}}{%
4(r_{+}^{1})^{2}}v\right) \right] ,w^{\prime \prime }=\arctan \left[ -\exp
\left( \frac{-r_{+}^{1}+r_{-}^{1}}{4(r_{+}^{1})^{2}}w\right) \right]
\]%
and multiplying the last term on the conformal factor $\Omega .$ We obtain
\begin{eqnarray}
\delta s^{2} &=&\Omega \left( \xi ,\theta ,\varphi \right) [-r^{2}qd\theta
^{2}-\eta _{3}(r,\theta ,\varphi )r^{2}\sin ^{2}\theta \delta \varphi ^{2}]
\label{el2a} \\
&&+64\frac{(r_{+}^{1})^{4}}{(r_{+}^{1}-r_{-}^{1})^{2}}\left( 1-\frac{2m}{r}+%
\frac{\varepsilon }{r^{2}}\Phi \left( \varphi \right) \right) \delta
v^{\prime \prime }\delta w^{\prime \prime },  \nonumber
\end{eqnarray}%
where $r$ is defined implicitly by
\[
\tan v^{\prime \prime }\tan w^{\prime \prime }=-\exp \left[ \frac{%
r_{+}^{1}-r_{-}^{1}}{2(r_{+}^{1})^{2}}r\right] \sqrt{\frac{r-r_{+}^{1}}{%
(r-r_{-}^{1})^{\chi }}},\chi =\left( \frac{r_{+}^{1}}{r_{-}^{1}}\right) ^{2}.
\]%
As particular cases, we may consider such $\eta _{3}\left( r,\theta ,\varphi
\right) $ as the condition of vanishing of the metric coefficient before $%
\delta v^{\prime \prime }\delta w^{\prime \prime }$ will describe a horizon
parametrized by a resolution ellipsoid hypersurface.

The maximal \ extension of the Schwarzschild metric polarized by a conformal
factor $\Omega $ and a function $\eta _{3}$ and deformed by a small
parameter $\varepsilon $ (for ellipsoid configurations treated as the
eccentricity), \ i. e. \ of the metric (\ref{sch}), is defined by taking (%
\ref{el2a}) as the metric on the maximal manifold on which this metric is of
smoothly class $C^{2}.$ The Penrose diagram of this static but locally
anisotropic space--time, for any fixed angular coordinates $\varphi _{0}$ is
similar to the Reissner--Nordstrom solution, for the case $e^{2}=\varepsilon
$ and $e^{2}<m^{2}$ (see, for instance, Ref. \cite{haw})). \ This way we
define a locally anisotropic conformal alternative to the locally
anisotropic space--time constructed in Ref. \cite{vo}. There is an infinite
number of asymptotically flat regions of type I, connected by intermediate
regions II and III, where there is still an irremovable singularity at $r=0$
for every region III. It is ''possible'' to travel from a region I to
another ones by passing through the 'wormholes' made by anisotropic
deformations (ellipsoid off--diagonality of metrics, or anholonomy) like in
the Reissner--Nordstrom universe because $\sqrt{\varepsilon }$ may model an
effective electric charge. We can not turn back in a such anisotropic travel.

Finally, in this section, we note that the metric (\ref{el2a}) $\ $\ is
analytic every were except at $r=r_{-}^{1}.$ We may eliminate this
coordinate degeneration by introducing another new coordinates%
\[
v^{\prime \prime \prime }=\arctan \left[ \exp \left( \frac{%
r_{+}^{1}-r_{-}^{1}}{2n_{0}(r_{+}^{1})^{2}}v\right) \right] ,w^{\prime
\prime \prime }=\arctan \left[ -\exp \left( \frac{-r_{+}^{1}+r_{-}^{1}}{%
2n_{0}(r_{+}^{1})^{2}}w\right) \right] ,
\]%
where the integer $n_{0}\geq (r_{+}^{1})^{2}/(r_{-}^{1})^{2}.$ In \ these
coordinates, the metric is analytic every were except at $r=r_{+}^{1}$ where
it is degenerate.$\,$\ This way the space--time manifold can be covered by
an analytic atlas by using coordinate carts defined by $(v^{\prime \prime
},w^{\prime \prime },\theta ,\varphi )$ and $(v^{\prime \prime \prime
},w^{\prime \prime \prime },\theta ,\varphi ).$ For maximally analytic
extensions of anisotropic metrics we must perform also extensions to
non--singular values of polarization coefficients.

\section{Conformally Anisotropic Geodesics}

The analysis of small deformations of geodesics, linear on $\varepsilon ,$
of the metric (\ref{sch1}) with the data (\ref{data}) is similar to to that
presented in Ref. \cite{vo} for the anisotropic metrics with trivial
conformal factor. We introduce the effective Lagrangian (see, for instance,
Ref. \cite{chan})%
\begin{eqnarray}
2L &=&g_{\alpha \beta }\frac{\delta u^{\alpha }}{ds}\frac{\delta u^{\beta }}{%
ds}=\Omega \left( r,\varphi \right) \times \{-\left( 1-\frac{2m}{r}+\frac{%
\varepsilon }{r^{2}}\right) ^{-1}\left( \frac{dr}{ds}\right) ^{2}
\label{lagr} \\
&&-r^{2}q(r)\left( \frac{d\theta }{ds}\right) ^{2}-\eta _{3}(r,\theta
,\varphi )r^{2}\sin ^{2}\theta \left( \frac{d\varphi }{ds}+\frac{\varepsilon
}{\eta _{3}^{\ast }}\eta _{3}^{\bullet }\frac{dr}{ds}\right) ^{2}  \nonumber
\\
&&+\left( 1-\frac{2m}{r}+\frac{\varepsilon }{r^{2}}\right) [\frac{dt}{ds}%
+\varepsilon \left( n_{1[1]}(r)+n_{1[2]}(r)\int \eta _{3}(r,\varphi
)d\varphi \right) \frac{dr}{ds}]^{2}\},  \nonumber
\end{eqnarray}%
for $\eta _{3}^{\bullet }=\partial \eta _{3}/\partial r$ and $r=r(\xi ).$

The corresponding Euler--Lagrange equations,
\[
\frac{d}{ds}\frac{\partial L}{\partial \frac{\delta u^{\alpha }}{ds}}-\frac{%
\partial L}{\partial u^{\alpha }}=0
\]%
are%
\begin{eqnarray}
&&\frac{d}{ds}\left[ -r^{2}q(r)\Omega \frac{d\theta }{ds}\right] =-\eta
_{3}r^{2}\sin \theta \cos \theta \left( \frac{d\varphi }{ds}\right) ^{2}+
\label{lag1} \\
&&\varepsilon \eta _{3}r^{2}\sin \theta \cos \theta \left[ (1-\frac{2m}{r}+%
\frac{\varepsilon }{r^{2}})\left( \frac{dt}{ds}\right) ^{2}-\left( \frac{%
d\xi }{ds}\right) ^{2}-2\frac{\eta _{3}^{\bullet }}{\eta _{3}^{\ast }}\frac{%
d\xi }{ds}\frac{d\varphi }{ds}-r^{2}q\left( \frac{d\theta }{ds}\right) ^{2}%
\right] ,  \nonumber \\
&&\frac{d}{ds}\left[ -\eta _{3}r^{2}(\frac{d\varphi }{ds}+\varepsilon \frac{%
\eta _{3}^{\bullet }}{\eta _{3}^{\ast }}\frac{d\xi }{ds})\Omega \right]
=-\eta _{3}^{\ast }\frac{r^{2}}{2}\sin ^{2}\theta \left( \frac{d\varphi }{ds}%
\right) ^{2}+  \label{lag2} \\
&&\varepsilon \{\left( 1-\frac{2m}{r}\right) n_{1[2]}(\xi )\eta _{3}\frac{%
d\xi }{ds}\frac{d\theta }{ds}+\frac{\eta _{3}^{\ast }}{2}r^{2}\sin
^{2}\theta \lbrack \left( 1-\frac{2m}{r}\right) \left( \frac{dt}{ds}\right)
^{2}-\left( \frac{d\xi }{ds}\right) ^{2}  \nonumber \\
&&-r^{2}q\left( \frac{d\theta }{ds}\right) ^{2}-\eta _{3}(1+r^{2}\sin
^{2}\theta )\left( \frac{d\varphi }{ds}\right) ^{2}-(\frac{\eta
_{3}^{\bullet }}{\eta _{3}^{\ast }}-\frac{\eta _{3}}{\eta _{3}^{\ast }}%
\left( \frac{\eta _{3}^{\bullet }}{\eta _{3}^{\ast }}\right) ^{\ast })\frac{%
d\xi }{ds}\frac{d\varphi }{ds}]\},  \nonumber
\end{eqnarray}%
and%
\begin{equation}
\frac{d}{ds}\left\{ (1-\frac{2m}{r}+\frac{\varepsilon }{r^{2}})\Omega \left[
\frac{dt}{ds}+\varepsilon \left( n_{1[1]}+n_{1[2]}\int \eta _{3}d\varphi
\right) \frac{d\xi }{ds}\right] \right\} =0,  \label{lag3}
\end{equation}%
where we have omitted the variations for $d\xi /ds$ which may be found from (%
\ref{lagr}) and considered that the polarization $\eta _{3}$ is taken as to
have $h_{3}$ depending only on variables $r$ and $\varphi .$ The system of
equations (\ref{lag1})--(\ref{lag3}) transform into the usual system of
geodesic equations for the Schwarzschild space--time if $\varepsilon
\rightarrow 0$ and $\eta _{3}\rightarrow 1$ which can be solved exactly \cite%
{chan}. For nontrivial values of the parameter $\varepsilon ,$ conformal
factor $\Omega $ and polarization $\eta _{3},$ it is a cumbersome task to
construct explicit solutions. We do not solve this problem in the present
work. We conclude only that from the equation (\ref{lag3}) one follows the
existence of an energy like integral of motion, $E=E_{0}+$ $\varepsilon
E_{1},$ with%
\begin{eqnarray*}
E_{0} &=&\left( 1-\frac{2m}{r}\right) \frac{dt}{ds} \\
E_{1} &=&\left( \frac{1}{r^{2}}+r^{2}\eta _{3}\right) \frac{dt}{ds}+\left( 1-%
\frac{2m}{r}\right) \left( n_{1[1]}+n_{1[2]}\int \eta _{3}d\varphi \right)
\frac{d\xi }{ds}.
\end{eqnarray*}

The anisotropic conformal deformations of the congruences of
Schwarzschild's \\ spa\-ce--ti\-me geodesics preserve the known
behaviour in the vicinity of the horizon hypersurface defined by
the condition of vanishing of the
coefficient $1-2m/r+\varepsilon \Phi \left( r,\varphi \right) /r^{2}$ in (%
\ref{el2a}). We can prove this by considering radial null
geodesics in the ''equatorial plane'' satisfying the condition
(\ref{lagr}) with $\theta =\pi /2,d\theta /ds=0,d^{2}\theta
/ds^{2}=0$ and $d\varphi /ds=0,$ when
\[
\frac{dr}{dt}=\pm \left( 1-\frac{2m}{r}+\frac{\varepsilon }{r^{2}}\right) %
\left[ 1+\varepsilon \left( n_{1[1]}+n_{1[2]}\int \eta _{3}d\varphi \right) %
\right] .
\]%
The integral of this equation is
\[
t=\pm r^{\Vert }+\varepsilon _{0}\int \left[ n_{1[1]}+n_{1[2]}\int \eta
_{3}d\varphi \right] dr
\]%
where the coordinate $r^{\Vert }$ is defined in equation (\ref{r1}). Even
the explicit form of the integral depends on the type of polarizations $\eta
_{3}(r,\theta ,\varphi )$ and $n_{1[1,2]}(r),$ which may result in small
deviations of the null--geodesics for certain prescribed class of
non--singular polarizations, we conclude that for an in--going null--ray the
coordinate time $t$ increases from $-\infty $ to $+\infty $ as $r$ decreases
from $+\infty $ to $r_{+}^{1},$ decreases from $+\infty $ to $-\infty $ as $%
r $ further decreases from $r_{+}^{1}$ to $r_{-}^{1},$ and increases again
from $-\infty $ to a finite limit as $r$ decreses from $r_{-}^{1}$ to zero.
We have a behaviour very similar to that for the Reissner--Nordstrom
solution but with some additional anisotropic contributions being
proportional to $\varepsilon .$ The conformal factor $\Omega $ contributes
indirectly via modification of the coefficients $n_{1[1,2]}$ in the
solutions. We also note that as $dt/ds$ tends to $+\infty $ for $%
r\rightarrow r_{+}^{1}+0$ and to $-\infty $ as $r_{-}+0,$ any radiation
received from infinity appear to be infinitely red--shifted at the crossing
of the event horizon and infinitely blue--shifted at the crossing of the
Cauchy horizon.

Following the mentioned properties of null--geodesics, we conclude that the
metric (\ref{sch}) (equivalently, (\ref{sch1})) with the data (\ref{data})
and it maximal analytic extension (\ref{el2a}) define an alternative (to
that constructed in Ref. \cite{vo}) black hole static solution which is
obtained by an anisotropic small (on $\varepsilon )$ deformation of the
Schwarzchild solution (for a particular type of deformations the horizon of
such black holes is defined by ellipsoid hypersurfaces). We call such
objects as black conformal ellipsoids. They exist in the framework of
general relativity as certain vacuum solutions of the Einstein equations
defined by static off--diagonal metrics and associated anholonomic frames.
This property disinguishes them from similar configurations of
Reissner--Norstrom type (which are static electrovacuum solutions of the
Einstein--Maxwell equations) and of Kerr type rotating solutions, in the
last case also with ellipsoid horizons but parametrized by off--diagonal
vacuum metrics constructed for a spherical coordinate system which induces a
holonomic local frame.

\section{Conclusions}

In summary, we proved that there are alternative small static conformal
anisotropic deformations of the Schwarschild black hole solution (as
particular cases we can consider resolution ellipsoid lconfigurations) which
have a horizon and geodesic congruence being slightly deformed from the
spherical symmetric case. There are possible different types of
parametrization of off--diagonal metrics defining the exact solutions of
vacuum Einstein equations which posses ellipsoid symmetries as it was
constructed in Refs. \cite{v,vth,vo}.

We can generate static ellipsoid black hole configurations with trivial, or
non--trivial conformal factors. The corresponding metrics admit maximial
analytic extensions and with the Penrose diagrams being similar to the
Reissner--Nordstrom solution. There is a similarity with the
Reissner--Nordstrom metric if the parameter of a conformal ellipsoid
deformation (the eccentricity) is treated as an effective electromagnetic
charge induced by off--diagonal vacuum gravitational interactions.

As we concluded in Ref. \cite{vo} the static ellipsoid black holes posses
spherical topology and satisfy the principle of topological censorship \cite%
{haw1}. The anisotropic conformal transforms, emphasized in this paper, do
not violate such principles. The existence of alternative classes of black
ellipsoid solutions does not contradict the black hole uniqueness theorems %
\cite{ut} which were formulated for spherically symmetric solutions. At
asymptotics, at least for a very small eccentricity, the anisotropic metrics
transform into the usual Schwarzschild one.


\subsection*{Acknowledgements}

~~ The S. V. work is supported by a NATO/Portugal fellowship at CENTRA,
Instituto Superior Tecnico, Lisbon.



\end{document}